\documentclass[prl,showpacs,twocolumn]{revtex4}
\usepackage{amsfonts}
\usepackage{amsmath}
\usepackage{amssymb}
\usepackage{graphicx}

\begin{document}

\preprint{}
\title[MEMT]{Magneto-electric momentum transfer to atoms and molecules}
\author{G.L.J.A. Rikken\medskip }
\affiliation{Laboratoire National des Champs Magn\'{e}tiques Intenses\\
UPR3228 CNRS/INSA/UJF/UPS, Toulouse \& Grenoble, France.}
\author{B.A. van Tiggelen}
\affiliation{Universit\'{e} Grenoble 1 and CNRS, LPMMC UMR 5493, 38042 Grenoble, France}

\begin{abstract}
We report the first observation of mechanical momentum transferred to atoms
and molecules upon application of crossed electric and magnetic fields. We
identify this momentum as the microscopic analogue of the classical Abraham
force. Several predictions of additional magneto-electrically induced
mechanical momentum are addressed. One of them, proposed to result from the
interaction with the quantum vacuum, is experimentally refuted, others are
found to be currently below experimental detection.
\end{abstract}

\volumeyear{year}
\volumenumber{number}
\issuenumber{number}
\eid{identifier}
\date{\today }
\startpage{101}
\endpage{102}
\maketitle

It has been shown that in crossed electric and magnetic field $\mathbf{E}$
and $\mathbf{B}$, the optical and electrical properties of matter become
anisotropic along the axis\ $\mathbf{E\times B}$ \cite{Rikken1} \cite%
{Rikken2} \cite{Rikken3}. As this anisotropy manifests itself in the
dispersion law and thus in the momentum of the photons and the charge
carriers respectively, one may wonder whether a similar anisotropy can exist
in the mechanical momentum of particles in crossed fields.

Invariance under time, charge and parity reversal straightforwardly shows it
to be symmetry allowed for a particle to acquire a mechanical momentum $%
\mathbf{p}$ upon applying a crossed electric and magnetic field $\mathbf{E}$
and $\mathbf{B}$
\begin{equation}
\mathbf{p}=a\mathbf{E\times B}  \label{vpropB}
\end{equation}%
If we assume the particles to be in a gaseous phase in a container, and the
collisions between the particles to be elastic, the momentum imparted to
each of the particles by the application of the fields will be conserved
within the gas as a whole, and ultimately transferred to the wall
perpendicular to $\mathbf{E\times B}$. If we apply a harmonically
oscillating electric field $\mathbf{E}(t)=\mathbf{E}\sin \omega t$ and a
static magnetic field $\mathbf{B}$\textbf{,} each particle will contribute a
force on this wall given by%
\begin{equation}
\mathbf{F}_{p}=\frac{d\mathbf{p}}{dt}=a\omega \mathbf{E\times B}
\label{momentumchange}
\end{equation}%
Such a force exerted on individual atoms would be the microscopic equivalent
of the so-called Abraham force density which was first formulated for
macroscopic media, and has been the subject of a long-standing controversy
\cite{Brevik1,nelson,Pfeiffer,Barnett}. In the so-called Abraham version one
finds a macroscopic force density (in SI units) $\mathbf{f}_{A}=\varepsilon
_{0}\left( \varepsilon _{r}-1/\mu _{r}\right) \mathbf{E}\times \mathbf{B}$
\cite{LandauLifshitz}, whereas in the Minkowski version $\mathbf{f}%
_{M}=\varepsilon _{0}\left( \varepsilon _{r}-1\right) \mathbf{E}\times
\mathbf{B}$ ($\varepsilon _{r}$ and $\mu _{r}$ are the relative dielectric
permittivity and magnetic permeability respectively). This can be compared
to the quantum-mechanical conserved pseudo-momentum of a neutral atom in a
homogeneous magnetic field, $\mathbf{K}=\sum_{i}m_{i}\mathbf{\dot{r}}%
_{i}+\sum_{i}q_{i}\mathbf{B}\times \mathbf{r}_{i}$ \cite{Kawka}. An
additional electric field creates a finite polarization $<\mathbf{P}%
>=\left\langle \sum_{i}q_{i}\mathbf{r}_{i}\right\rangle =\alpha \mathbf{E}$ (%
$\alpha $ is the static electric polarizability of the particle with SI unit
$Cm^{2}/V$) in the ground state so that $0=\mathbf{\dot{K}}=\sum_{i}m_{i}%
\mathbf{\ddot{r}}_{i}-\alpha \partial _{t}\mathbf{E}\times \mathbf{B}$. This
would lead to a force density $\mathbf{f}=N\alpha \partial _{t}\mathbf{E}%
\times \mathbf{B}$ (where $N$ is the particle density) and which is
consistent with the Minkowski version, since $\epsilon _{0}(\epsilon
_{r}-1)=N\alpha $, and we deduce $a=\alpha $ in Eq.(\ref{vpropB}). Note that
the pseudo-momentum in this model equals neither the \emph{conjugated}
momentum $\mathbf{P}=\sum_{i}m_{i}\mathbf{\dot{r}}_{i}+\frac{1}{2}%
\sum_{i}q_{i}\mathbf{B}\times \mathbf{r}_{i}$, nor the kinetic momentum $%
\mathbf{P}_{\mathrm{kin}}=\sum_{i}m_{i}\mathbf{\dot{r}}_{i}$. Both were
proposed by Barnett \cite{Barnett} to solve the Abraham-Minkowski
controversy.

The observation of the Abraham force due to a crossed oscillating electric
field and a static magnetic field was reported by James \cite{James} and by
Walker et al \cite{Walker} \cite{Walker2} in solid dielectrics. It should be
noted that the Abraham force due to a static electric field and an
oscillating magnetic field was reported \textit{not} to be observed, against
all expectation \cite{Walker3,Walker4,Walker5}. For a discussion of these
and related experiments, see \cite{Brevik1} and \cite{Pfeiffer}.

Feigel was the first to consider the interaction of a macroscopic
magneto-electric material with the quantum vacuum \cite{Feigel}. The
so-called Feigel effect implies that momentum from the vacuum fluctuations
can be transferred to matter by the intermediary of the optical
magneto-electric anisotropy and that therefore an QED contribution exists to
the classical Abraham force, corresponding to a 'Feigel' momentum $p_{F}$:%
\begin{equation}
p_{F}=\frac{1}{32N\pi ^{2}}\Delta n_{MEA}\hbar \left( \frac{\omega _{c}}{c}%
\right) ^{4}  \label{FeigelVelocity}
\end{equation}%
where $\Delta n_{MEA}\equiv \chi _{MEA}EB$ is the magneto-electric optical
anisotropy \cite{Rikken1} \cite{Rikken2}. In order to avoid the notorious UV
catastrophe, Feigel was obliged to introduce an empirical cut-off frequency $%
\omega _{c}$\ for the material's response. Particularly this cut-off
procedure was contested by several groups, since it is widely believed that
the UV catastrophe should somehow be absorbed in the parameter values
attributed to bulk media \cite{milton}. It was shown \cite%
{VanTiggelen1,VanTiggelen2,BirkelandBrevik} that such a transfer would then
only occur in a geometry of finite size, similar to that of the Casimir
effect, albeit with much smaller values than obtained by Feigel. Obukhov and
Hehl \cite{Obukhov2} also argued that no net momentum transfer from vacuum
fluctuations to bulk media can exist. However, very recently, Croze has
forwarded new theoretical support in favour of Feigel's claim \cite{Croze},
correcting in the process a minor numerical error. Kawka and Van Tiggelen
have proposed a nonrelativistic quantum theory of a harmonic oscillator in
crossed electric and magnetic fields \cite{Kawka}, in which the UV
catastrophe was shown to be absorbed in a mass renormalisation of the
oscillator. Applying this model to a hydrogen atom predicts a reduction of $%
a $ by 2\%.

For the ratio between the Feigel and Abraham momenta we find%
\begin{equation}
\frac{p_{F}}{p_{A}}=\frac{\pi ^{2}}{2}\frac{\chi _{MEA}}{N\alpha }\left(
\frac{1}{\lambda _{c}}\right) ^{4}\hbar   \label{RatioVelocities}
\end{equation}%
where $\lambda _{c}=2\pi c/\omega _{c}$. Feigel proposed $\lambda _{c}=0,1\
nm$ as the limit of the matter response to the vacuum fluctuations. Using
the experimental results of Roth and Rikken \cite{Rikken1} for large
organometallic molecules, Croze predicts $p_{F}/p_{A}\approx 7$, i.e. the
magneto-electrically induced particle momentum would be dominated by the
contribution from the quantum vacuum.


\begin{figure}[htb]
\begin{center}\label{awesome_image}
\leavevmode
\includegraphics[width=0.7\linewidth,keepaspectratio]{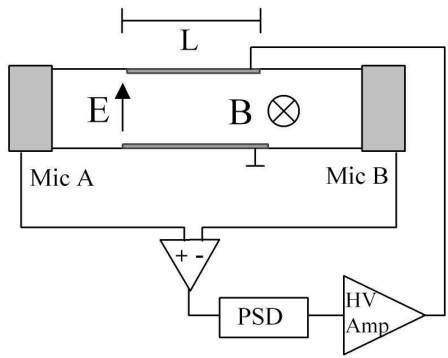}
\end{center}
\caption{Schematic setup of the experiment.}

\end{figure}

Clearly the magneto-electrical momentum, in spite of its long history, still
poses fundamental problems. In this Letter we will describe a new method to
accurately measure the momentum transferred to atoms or molecules in crossed
oscillating electric fields and static magnetic fields. We experimentally
confirm\ the prediction of the classical Abraham force. More specifically,
we do not observe any deviations from the Abraham prediction for media where
the predicted contribution from the Feigel effect should be observable.

Since the work by James and by Walker \textit{et al}, no new experiments to
measure the Abraham force have been reported. Very recently, a proposition
was made to measure it at optical frequencies using whispering gallery modes
\cite{BrevikEllingsen}. The method used here measures directly the pressure
exerted by an atomic or molecular gas on the wall of the container if it is
exposed to crossed electric magnetic and magnetic fields, $E\widehat{\mathbf{%
x}}\sin \omega t$ and $B\widehat{\mathbf{y}}$ respectively. If we define the
effective length of the $\mathbf{E\times B}$ region as $L=\int E(z)B(z)dz/EB$%
,\ the momentum change due to the Abraham force exerts an oscillating
pressure $P$ on the wall perpendicular to $\mathbf{E\times B}$ given by
\begin{equation}
P(t)=\alpha \omega NLEB\cos \omega t  \label{pressure}
\end{equation}%
Such a pressure can be detected by a microphone located at the wall. By
tuning $\omega $ to a longitudinal acoustic resonance of the system, the
pressure can by multiplied by the Q factor of the resonance. Using values of
$N=2,7.10^{25}m^{-3}$ (1 bar ideal gas), $E=10^{5}V/m$, $B=1\;T$, $\omega
=3.10^{4}s^{-1},L=2\;cm$, $Q=10$ and $\alpha =$ $2,2.10^{-41}$ $Cm^{2}/V$\
we find $P=4.10^{-7}Pa$ and a velocity of \ $0,3$ $nm/s$. (values for He,
\cite{Handbook}). The typical sensitivity of an electret microphone is S =10
mV/Pa, so microphone signal voltages of around 5 nV can be expected which
are within experimental reach when using phase sensitive detection (PSD).
Figure 1 shows schematically the setup used. It consists of a 3 mm diameter,
5 cm long glass tube, with commercial electret microphones butt coupled to
its ends, carefully shielded in thick-walled copper housings. The electric
field was supplied by a high voltage amplifier (HV amp), generating voltages
up to 1000 V, and the magnetic field was provided by an electromagnet, with
fields up to 1,5 T. The Q factor was determined from the acoustic resonance
lineshape. The systematic inaccuracy of our setup is estimated to be 3 \%,
mostly due to the inaccuracy of the microphone sensitivity calibration.


\begin{figure}[htb]
\begin{center}
\leavevmode
\includegraphics[width=0.7\linewidth]{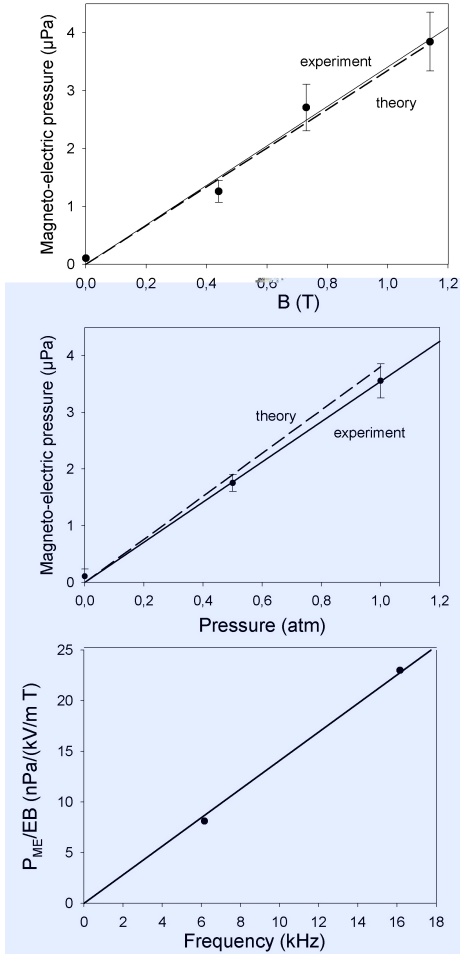}
\end{center}
\caption{Magneto-electric acoustic pressure observed in nitrogen gas.
Top panel; 6,15 kHz, 1 atm, E = 370 kV/m, Middle panel; 6,15 kHz, 1 atm,
B = 1,14 T. Bottom panel; 1 atm. Solid lines are linear fits to the data,
dashed lines theoretical predictions.}
\label{Figure2}
\end{figure}

\begin{figure}[htb]
\begin{center}
\leavevmode
\includegraphics[width=0.7\linewidth]{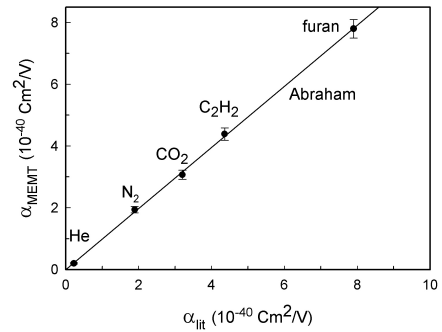}
\end{center}
\caption{Polarizability for different atoms and molecules as deduced
from magneto-electrically generated pressure, as a function of the literature values.
Solid line corresponds to the classical Abraham force prediction.}
\label{Figure3}
\end{figure}

Typical results for nitrogen gas are shown in Figure \ref{Figure2},
confirming the linear dependencies of the magneto-electrically induced
pressure on magnetic field strength, gas pressure ($\propto $ particle
density) and electric field oscillation frequency. The linear dependence on
electric field strength is intrinsic because of the phase sensitive
detection of the pressure signal. The dashed lines in the two top panels are
the theoretical predictions, based on Eq. \ref{pressure}. Within the
experimental accuracy, the experimental results agree with the theory. The
slope in the bottom panel of Figure \ref{Figure2} allows to determine $%
\alpha $, using Eq. \ref{pressure}. Figure \ref{Figure3} shows the results
for $\alpha $\ obtained this way for several gazes, as a function of the
literature value for $\alpha $. All gazes were measured at room temperature
and atmospheric pressure, except furan, which was measured at its room
temperature vapor pressure. Table 1 summarizes these results, and shows also
the calculated contribution of the Feigel momentum, expressed as a fraction
of the Abraham momentum, and based on experimental or theoretical values for
$\chi _{MEA}$. Only the value for nitrogen is experimental \cite{Robillard},
but it is in good agreement with the calculated value \cite{RizzoJones},
giving confidence in the other values calculated by the same authors. For
the two molecules in the table with the highest magneto-electric anisotropy,
the predicted contributions for the Feigel effect are much larger than the
experimental uncertainties on $\alpha _{MEMT}$, up to 7,5 times for furan.
As $\alpha _{MEMT}$ and $\alpha _{LIT}$ agree within the experimental
uncertainties, we conclude from these results that the prediction for the
Feigel effect as expressed by Eq. \ref{FeigelVelocity} is not observed. Note
that the Feigel prediction contains one adjustable parameter, the response
cutoff wavelength $\lambda _{c}$, and that increasing its value to 0,17 nm
decreases the prediction of the Feigel momentum contribution to below our
experimental uncertainty. However, strong magneto-electric anisotropy was
still reported at 0,16 nm wavelength \cite{Kubota}, the shortest wavelength
at which its observation was ever attempted. Our experimental results
therefore unambiguously contradict Feigel's prediction. Recent theoretical
work on simple models suggests that $\chi _{MEA}$ decays algebraically as $%
\omega ^{-2}$ at high frequencies, in much the same way as the dynamic
electrical polarizability \cite{Babington}. This makes the UV catastrophe in
the macroscopic description as proposed by Feigel, unavoidable and
unrepairable.

In a QED version of the Feigel effect by Kawka and Van Tiggelen \cite{Kawka}%
, this UV catastrophe was removed by mass regularization. Our current
experimental accuracy does not allow to make quantitative statements
concerning this prediction, but our setup could be improved to attain the
1\% accuracy estimated to be necessary for the observation of this
regularization. We hope that this perspective will stimulate realistic
calculations of this regularization, beyond the harmonic oscillator
approximation and in a relativistic context.

In order to make a contribution to the Abraham-Minkowski debate, our
experiment would have detect the difference between $1$ and $1/\mu _{r}$.
The gas with the largest $\mu _{r}$ to our knowledge is oxygen, with $\mu
_{r}-1=3,4\ 10^{-3}$ at room temperature and 1 atm. \cite{Handbook}.
Attaining such a precision is a considerable experimental challenge, but
going to lower temperatures or higher pressures could increase $\mu _{r}-1$
to accessible values.

In summary, we have reported the first observation of \ mechanical momentum
transferred to atoms and molecules by applying crossed time-varying electric
fields and static magnetic fields. We quantitatively identify this momentum
as the microscopic analogue of the classical Abraham force. We exclude the
existence of additional magneto-electrically transferred momentum, as
proposed by Feigel to result from the optical magneto-electric anisotropy
interacting with the quantum vacuum fluctuations. Other predictions for
additional contributions to the Abraham force are currently beyond our
experimental resolution, but the new method described in this Letter has
potential to successfully address these issues.


\begin{table}[tbp] \centering%
\begin{tabular}{lcccc}
{gas} & 
$ 
\begin{array}{c}
{\alpha }_{MEMT} \\
(10^{-40}\frac{Cm^{2}}{V})%
\end{array} $&
$\begin{array}{c}
{\alpha }_{LIT} \\
(10^{-40}\frac{Cm^{2}}{V})
\end{array}  $ &
$\begin{array}{c}
{\chi }_{MEA} \\
(10^{-22}\frac{m}{VT})
\end{array} $
 & ${p}_{F}{/p}_{A}\medskip $
 \\
He & $ 0,20\pm 10\% $ & $0,22$ {\cite{Handbook}} & $0,017 $ {\cite{RizzoJones}}
& $1,5\% $ \\
N$_{2}$ & $1,9 \pm 5\% $& $1,89$ {\cite{Handbook}} & $0,47$ {\cite{Robillard}}
& $4,8\%$ \\
C$_{2}$H$_{2}$ & $4,4\pm 5\% $ & $4,4$ {\cite{Handbook}} & $3,7$ {\cite
{RizzoJones}} & $16\% $\\ 
furan & $7,8 \pm 4\% $& $7,9$ {\cite{Kamada}} & $12$ {\cite{RizzoPrivate}} &
$29\%$
\end{tabular}%
\caption{Polarizabilities deduced from magneto-electric
momentum transfer, the corresponding literature values, the experimental
or calculated magneto-electric anisotropy,  and the calculated ratio of Feigel and
Abraham momenta for the gazes studied.\label{key}}%
\end{table}%

This work was supported by the ANR contract PHOTONIMPULS
ANR-09-BLAN-0088-01. We gratefully acknowledge helpful discussions and
calculations by Antonio Rizzo.

\end{document}